\documentclass[fleqn,usenatbib]{mnras}
\usepackage{mathptmx}

\usepackage[T1]{fontenc}
\usepackage{ae,aecompl}

\usepackage{graphicx}	
\usepackage{amsmath}	
\usepackage{amssymb}	
\usepackage{txfonts}
\usepackage{multirow}
\usepackage{mathrsfs}
\usepackage{natbib}

\title[The properties of the quasars astrometric solution]{The properties of the quasars astrometric solution in Gaia DR2}

\author[S. L. Liao et al.]{
Shilong Liao,$^{1,3}$\thanks{E-mail: shilongliao@shao.ac.cn (SLiao)}
Beatrice Bucciarelli,$^{2}$
Zhaoxiang Qi,$^{1,3}$
Sufen Guo,$^{1,3}$
Zihuang Cao$^{4}$\\
\newauthor{
and Zhenghong Tang$^{1,3}$}
\\
$^{1}$Shanghai Astronomical Observatory, Chinese Academy of Sciences, 80 Nandan Road, Shanghai 200030, P.R.China\\
$^{2}$INAF - Osservatorio Astrofisico di Torino, 20 Strada Osservatorio, Pino Torinese (TO) 10025, Italy\\
$^{3}$University of Chinese Academy of Sciences, No.19(A) Yuquan Road, Shijingshan District, Beijing 100049, P.R.China\\
$^{4}$National Astronomical Observatory, Chinese Academy of Sciences, 20A Datun Road, Chaoyang District, Beijing 100012, P.R.China 
}

\date{Accepted XXX. Received YYY; in original form ZZZ}

\pubyear{2015}

\begin{document}
\label{firstpage}
\pagerange{\pageref{firstpage}--\pageref{lastpage}}
\maketitle

\begin{abstract}
Gaia data release 2 (DR2) provides the best non-rotating optical frame aligned with the radio frame (ICRF) thanks to the inclusion of about half-million quasars in the 5-parameter astrometric solution. Given their crucial diagnostic role for characterizing the properties of the celestial reference frame, we aim to make an independent assessment of the astrometry of quasars in DR2. We cross-match with Gaia DR2 the quasars from LQAC3, SDSS and LAMOST, obtaining 208743 new sources (denominated as KQCG). With the quasars already identified in DR2, we estimate the global offsets of parallaxes and proper motions of different quasar subsets to check their astrometric consistency. The features of the proper motion field are analyzed by means of vectorial spherical harmonics (VSH); the scalar field of parallaxes is expanded into spherical harmonics to investigate their spatial correlation. We find a bias of $\sim$ $-0.030$ $mas$ in  KQCG parallaxes, and a bias of $-9.1$ $\mu as/yr$ in $\mu_{\alpha\ast}$; a $\sim$ +10 $\mu as/yr$ bias in $\mu_{\delta}$ consistently found in the entire quasar sample. The results of the VSH analysis on different subsets  indicate good  agreement between  them. The proper motion field exhibits a \((-6,-5,-5) \pm 1\) $\mu as/yr$ rotation in the northern hemisphere and a rotation of \((0,+1,+3) \pm 1 \) $\mu as /yr$ in the southern one. The spherical harmonics expansion of the parallax field reveals an angular scale of systematics of \(\approx \)18 degrees with a RMS amplitude of  13 \(\mu as\). The positional comparison shows that the axes of the current Gaia Celestial Reference Frame and the ICRF2 are aligned  within 30 \(\mu as\).
\end{abstract}




\section{Introduction}

The second Gaia data release (Gaia DR2) contains astrometric data for 1.693 billion sources from magnitude 3 to 21 based on the observations of the European Space Agency Gaia satellite during the 22-month period between 25 July 2014 and 23 May 2016 \citep{Lindegren2018Gaia}, hereafter cited as Gaia-DR2 Astrometry paper. Among all the sources with a full 5-parameter astrometric solution, DR2 provides more than 550 000 quasars, obtained from a positional cross-match with the ICRF3-prototype and the AllWISE AGN catalogues. These quasars are made to represent a kinematically non-rotating reference frame  (the celestial reference frame of Gaia, or Gaia-CRF2) in the optical domain \citep{Mignard2018Gaia}, hereafter denoted as Gaia-CRF2 paper.

Quasars (or QSOs) are extremely distant and small in apparent size. They are essential for absolute astrometry in the sense that they present no significant parallax or proper motion. Thus, they are ideal objects to investigate the properties of an astrometric solution. Besides the AllWISE AGN catalog \citep{Secrest2015Identification}, there are other catalogues that can enlarge the sample of quasars in Gaia DR2, such as the Large Quasar Astrometric Catalogue (LQAC) \citep{Souchay2015The},  the spectroscopically confirmed quasars in the SDSS-DR14 Quasar Catalog \citep{P2017The} and the spectroscopically confirmed quasars in LAMOST DR5 using LAMOST spectroscopic data \citep{Cui2012LAmost}; all these have been cross-matched with DR2 sources and collected in a comprehensive catalog named "Known Quasars Catalog for Gaia" (KQCG)  \citep{Liao2018KQCG}.  The aim of this paper is to make an independent assessment of the astrometry of quasars in DR2.

 After describing the quasar selection process in section 2, we address the global parallax and proper motion bias in section 3. In section 4,  we discuss the analysis of the proper motion field; the scalar spherical harmonics analysis of parallaxes is presented in section 5, and section 6 is devoted to the comparison between ICRF2 sources and their counterparts in Gaia DR2.The last section reports our conclusions. 

\section{Data Used}
Gaia DR2 includes  555934 quasars matched to the AllWISE AGN catalogue, plus 2820 sources matched to the ICRF3-prototype \citep{Lindegren2018Gaia,Mignard2018Gaia}. The union of these two sets makes  a total of 556869 sources, denoted as GCRF2. Among these, 485985 sources matched to the AllWISE AGN catalog are used to {\it define} a kinematically non-rotating reference frame, and are identified in the Gaia Archive by the field $frame\_rotator\_object\_type=3$ (Type3); whereas the 2820 sources matched to the ICRF3-prototype and used to align the GCRF2 axes with the radio ICRF are indicated by $frame\_rotator\_object\_type=2$ (Type2).

To maximize the size of our quasars sample, we cross-matched Gaia  DR2  with the compilation of SDSS-DR14Q, LQAC3 and LAMOST DR5 which are known to contain a huge number of reliable QSOs/AGNs. For the final selection, we adopt the joint conditions in Equation (14) of the Gaia DR2 Astrometric paper in order to reduce the risk of stellar contamination, as reported below:

   \begin{itemize}
      \item[(i)]   astrometric$\_$matched$\_$observations $\ge$ 8,
      \item[(ii)] astrometric$\_$params$\_$solved=31,
      \item[(iii)]  $\left|(\omega+0.029 mas)/\sigma_{\omega}\right|$<5,
      \item[(iv)] $(\mu_{\alpha^{\ast}}/\sigma_{\mu\alpha^{\ast}})^2+(\mu_{\delta}/\sigma_{\mu\delta})^2<25$,
      \item[(v)]  $\left|\sin b\right|$>0.1
     \item[(vi)] $\rho$<(2 arcsec)$\times$$\left|\sin b\right|$
   \end{itemize}
Where $\rho$ is the radius for the positional matching, $b$ is the Galactic latitude.

\begin{figure}
   \centering
   \includegraphics[width=6cm]{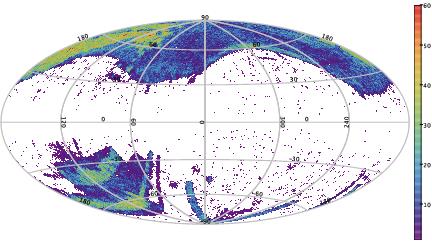}
      \caption{The sky distribution of KQCG. The map shows the sky density with each cell of approximately 0.84 $deg^{2}$, using the Hammer-Aitoff projection in Galactic coordinates with zero longitude at the centre and increasing longitude from right to left.}
         \label{Figkqcg-non-define}
   \end{figure}

\begin{figure}
   \centering
   \includegraphics[width=6cm]{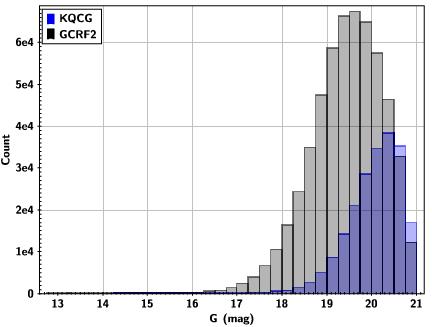}
      \caption{G magnitude distribution of the Gaia-CRF2 sources and the KQCG sources.}
         \label{FigGmag_kqcg}
   \end{figure}

With these precepts, we found 208743 new  quasars (KQCG) in Gaia DR2. There are about 87$\%$ of the quasars located in the northern hemisphere. The sky density distribution of  the KQCG catalog is depicted in Figure \ref{Figkqcg-non-define}; Figure \ref{FigGmag_kqcg} shows the histograms of the G-magnitude distribution for the GCRF2 and KQCG samples, indicating that our additional quasars populate the dimmer end, and most of them are fainter than $G=19$. 

   \begin{figure}
   \centering
   \includegraphics[width=6cm]{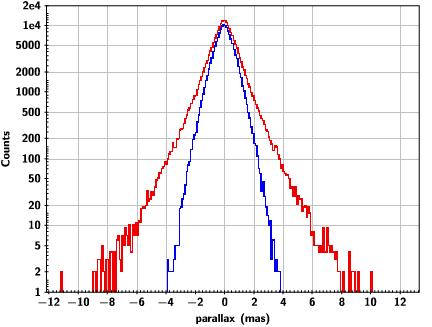}
      \caption{Parallaxe distribution for the KQCG quasars. Outer (red) curve is the whole KQCG sample; inner (blue) is the subsample of 143806 sources with $\sigma_{\omega}$<1 mas.}
         \label{FigParallax_kqcg}
   \end{figure}

   \begin{figure}
   \centering
   \includegraphics[width=6cm]{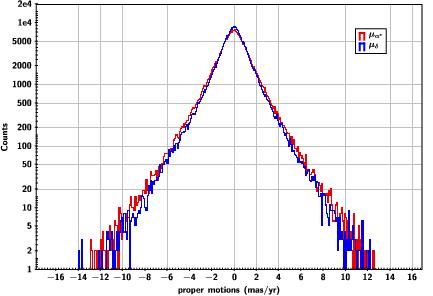}
      \caption{Proper motions distribution for the  KQCG. The red curve is the proper motions in right ascension $\mu_{\alpha\ast}$ and the blue is the proper motions in declination $\mu_{\delta}$.}
         \label{FigPM_kqcg}
   \end{figure}

\begin{figure}
\centering
\includegraphics[width=.30\textwidth]{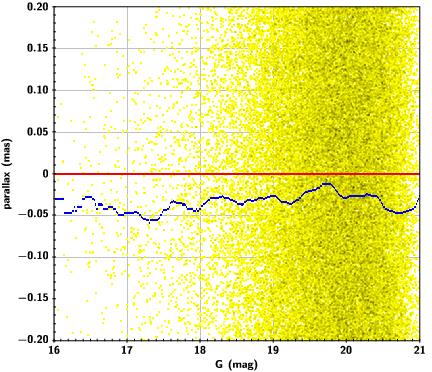}
\includegraphics[width=.30\textwidth]{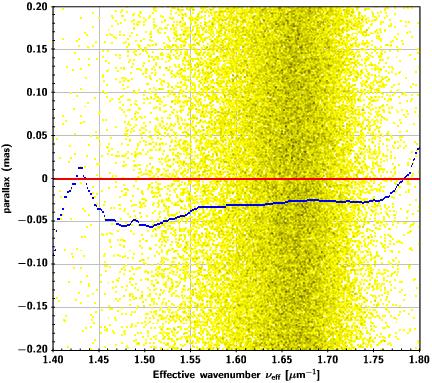}
      \caption{Parallaxes of the KQCG quasars plotted against Gaia G magnitude (top), colour (bottom). The yellow dots are the parallax data points, the blue lines are the parallax medians $\omega_{med}$ of each running-bin.}
         \label{FigParallax_bisa_kqcg}
\end{figure}

\section{Global bias}
\subsection{Parallax zero point}
  Figure \ref{FigParallax_kqcg} shows the distribution of parallaxes for the complete KQGC (red curve) and the high-precision subset (blue curve).  The  mean and median parallax of the whole sample are  $-0.0330$ $mas$ and $-0.0278$ $mas$, respectively; the corresponding values for the high-precision subset are $-0.0270$ $mas$ and $-0.0264$ $mas$. Table \ref{parallax_bias} gives the different averages calculated for each data sample. The weighted mean parallax is consistent between different subsets, setting to about $-0.029$ $mas$. However,  the mean parallax for $Type2$  is sensibly smaller, offsetting by  $0.02$ $mas$ from the other samples. Plots of parallax versus magnitude and effective wavenumber,  closely related to the source colour, are shown in Figure \ref{FigParallax_bisa_kqcg}, which reveals the presence of  trends in the  systematic parallax error, with an excursion of $\sim$0.020 mas over the range covered by the data.

\begin{table}
\centering
\caption{The mean and median parallax (in mas) of different quasar subsets. The formal parallax error is used as weight to calculate the weighted average.}
\label{parallax_bias}
\begin{tabular}{ccccc}
\hline
\hline
&&&&\\
Subset&N&Mean&Weighted Mean &Median\\
\hline
&&&&\\
KQCG                    & 208743             & -0.0330                           & -0.0291                            & -0.0278            \\
GCRF2                   & 556869             & -0.0308                           & -0.0292                           & -0.0287      \\
Type2                   & 2843               & -0.0511                          & -0.0382                           & -0.0351    \\
Type3                   & 485985             & -0.0284                           & -0.0283                            & -0.0281   \\
\hline
\end{tabular}
\end{table}

\subsection{Proper motion bias}
\label{pmbias}
\begin{figure*}
   \centering
   \includegraphics[width=4.15cm]{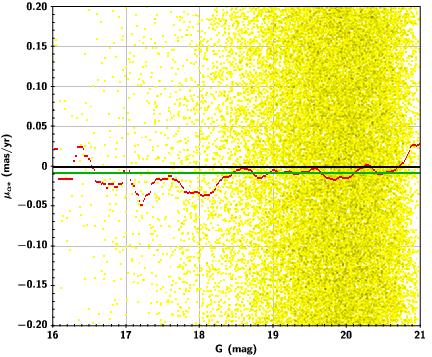}
   \includegraphics[width=4.10cm]{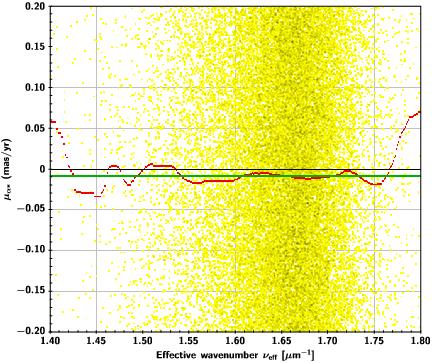}
   \includegraphics[width=4.15cm]{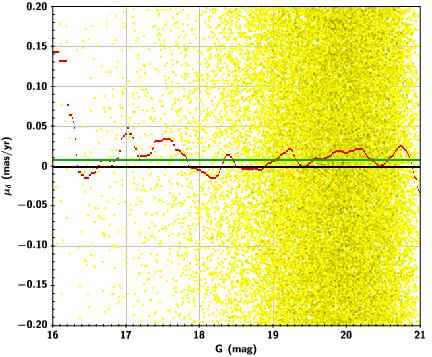}
   \includegraphics[width=4.10cm]{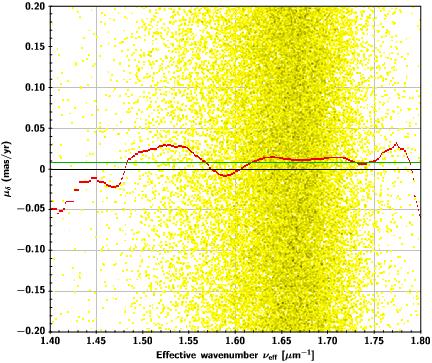}
      \caption{Proper motions of the KQCG  plotted against the Gaia G magnitude and colour (the first and second panel from the left are for $\mu_{\alpha\ast}$, and the third and fourth are for $\mu_{\delta}$). The yellow dots are the proper motion data. The green line is the mean proper motion, while the red lines are the proper motion medians of each running-bin.}
         \label{kqcgpm}
   \end{figure*}

\begin{figure*}
   \centering
   \includegraphics[width=4.15cm]{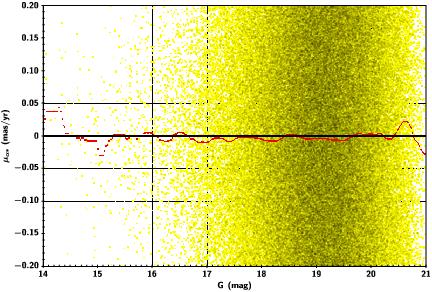}
   \includegraphics[width=4.2cm]{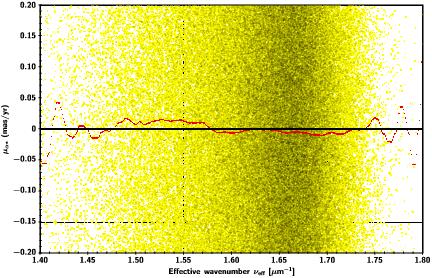}
   \includegraphics[width=4.2cm]{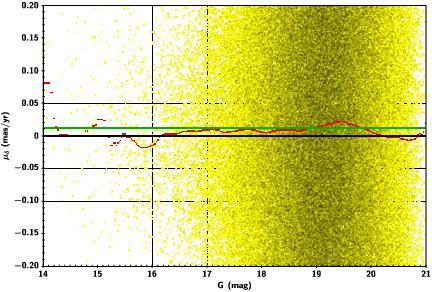}
   \includegraphics[width=4.3cm]{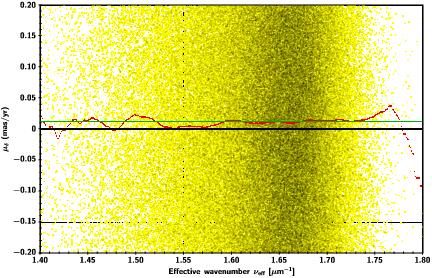}
      \caption{Proper motions of the GCRF2  plotted against the Gaia G magnitude and colour (the first and second panel from the left are for $\mu_{\alpha\ast}$, and the third and fourth are for $\mu_{\delta}$). The yellow dots are the proper motion data. The green line is the mean proper motion, while the red lines are the proper motion medians of each running-bin.}
         \label{gcrfpm}
   \end{figure*}

Besides parallaxes, the proper motions of quasars are also nominally zero (the Galactic acceleration effect is neglected here). Figure \ref{FigPM_kqcg} shows the distribution of the proper motion for the KQCG sample;  Table \ref{men_med_pm} gives the mean and median proper motion of the different subsets. For the GCRF2 sample we obtain $+1.8$ $\mu as/yr$ and $-1.5$ $\mu as/yr$ in $\mu_{\alpha\ast}$, which is near zero;  however, the  mean and median in $\mu_{\delta}$ raise to $+12.3$ $\mu as/yr$ and $+11.7$ $\mu as/yr$. For the KQCG sample, the corresponding values are $-8.7$ $\mu as/yr$ and $-7.5$ $\mu as/yr$ in $\mu_{\alpha\ast}$,  and +8.3 $\mu as/yr$ and $+11.4$ $\mu as/yr$ in $\mu_{\delta}$, respectively. Looking at the Type2 sample, we get  $+10$ $\mu as/yr$ in both components. If we take weighted averages based on formal errors, only the KQCG sample has a significant bias of about $-9.1$ $\mu as/yr$ in $\mu_{\alpha\ast}$, while there is a common bias of $+10$ $\mu as/yr$ in declination for  all subsets. The distribution of proper motion versus magnitude and effective wavenumber of  KQCG and GCRF2 are plotted in Figures \ref{kqcgpm} and \ref{gcrfpm}. In the second panel of  Figure \ref{gcrfpm}, the median proper motion of $\mu_{\alpha\ast}$ trends from positive to negative at the effective wavenumber $\nu_{eff}\sim$ 1.58 $\mu m^{-1}$ ; this result seems in agreement with the findings of  the Gaia-DR2 Astrometric paper, see their Figure 3. Interestingly, the KQCG sample does not clearly follow the same trend as function of the effective wavenumber, suggesting either a different correlation between magnitude and color for these quasars, or a more complex color dependence of the astrometric calibration for fainter objects.

\begin{table*}
\centering
\caption{The mean and median proper motion of different quasar subsets. The proper motion error is used as weight.}
\label{men_med_pm}
\begin{tabular}{cccccccc}
\hline
\hline
\multirow{2}{*}{Subset} & \multirow{2}{*}{N} & \multicolumn{3}{c}{$\mu_{\alpha\ast}$$(\mu as/yr)$} & \multicolumn{3}{c}{$\mu_{\delta}$$(\mu as/yr)$} \\
                        &                    & Mean             &Weighted Average              & Median  & Mean        &Weighted Average        & Median       \\
\hline
&&&&&\\
KQCG                    & 208743             & -8.7             &-9.1              & -7.5      & +8.3             &+11.1           & +11.4                         \\
GCRF2                   & 556869             & +1.8            &-0.7               & -1.5       &+12.2         & +12.3            & +11.7            \\
Type2                   & 2843                & +16.1         &+2.9           & +10.5                           & +19.3      &+14.7             & +8.1                          \\
Type3                   & 485985             & +0.3       &-1.3            & -1.4                            & +11.9       &+11.8        & +11.7                        \\
\hline
\end{tabular}
\end{table*}

\section{Analysis of the proper motion field}\label{pmvsh}
In this section, we perform the vector spherical harmonics (VSH) analysis of different quasar samples. 
The results of the VSH analysis are listed in Table \ref{vshresults}. After adding the KQCG sample to GCRF2 (KQCG plus GCRF2, denoted as KG), rotation and glide do not change very much between the harmonics of degree $l=1$ and $l=10$, and agree well with the results of the GCRF2 sample. Since the quasars in KQCG are mostly fainter than 19 and are not uniformly  distributed, we also compare two subsamples ($19\leq$ G< 20 and G$\geq$ 20) of KG and GCRF2. The results agree with each other, which indicates consistency between the astrometric solutions.

As pointed out in Section \ref{pmbias}, the median proper motion of $\mu_{\alpha\ast}$ trends from positive to negative at the effective wavenumber $\nu_{eff}\sim$ 1.58 $\mu m^{-1}$ for GCRF2 sample. The VSH analysis result shows that the two quasar subsets ($\nu_{eff}$ $\geq$ 1.58 and $\nu_{eff}$ < 1.58) have a similar glide but a very different rotation (mainly $x$ and $y$ components).

The glide results agree with different subsets with a typical glide value of $(-9, +5 ,+12)\pm1$ $\mu as/yr$.  If we subtract the global proper motion bias in both components before performing the VSH analysis, the typical glide is $(-9, +5 ,-2)\pm1$ $\mu as/yr$, see the rows marked with $\ast$ in Table \ref{vshresults}.

\begin{table*}
\centering
\caption{VSH analysis of the proper motion field of different quasar subsets in Gaia DR2. In the rows marked with $\ast$ the mean proper motion is subtracted before the VSH analysis is performed. All solutions are weighted. "-" means no estimation.}
\label{vshresults}
\begin{tabular}{ccccccccc}
\hline
\hline
& &  &  &  &&  &  &    \\
\multirow{2}{*}{\begin{tabular}[c]{@{}c@{}}Source\\ Subset\end{tabular}} & \multirow{2}{*}{$l_{max}$} & \multirow{2}{*}{N} & \multicolumn{3}{c}{Rotation[$\mu$as/yr]} & \multicolumn{3}{c}{Glide[$\mu$as/yr]} \\
 &  && x & y & z & x & y & z \\
\hline
& &  & & &  &  &  &    \\
 GCRF2& 5 &556869& -5.5$\pm$1.1 & -7.4$\pm$0.9 & 5.6$\pm$1.2 & -9.2$\pm$1.2 & 4.7$\pm$1.0 & 11.6$\pm$1.0 \\
$\ast$&5 &556869& -5.5$\pm$1.1 & -7.4$\pm$0.9 & 3.5$\pm$1.2 & -9.1$\pm$1.2 & 4.8$\pm$1.0 & -2.9$\pm$1.0 \\
19$\leq$G<20&5&257446&10.9$\pm$2.3 & 3.5$\pm$1.8 & 6.7$\pm$2.5 &-7.8$\pm$2.5  &1.9$\pm$2.0  &16.8$\pm$2.0    \\
G$\geq$20&5&148910&3.3$\pm$6.0 & 27.5$\pm$5.4 & 5.4$\pm$7.2 &-15.3$\pm$6.7  &12.6$\pm$5.7  &7.6$\pm$5.7 \\
& &  & & &  &  &  &    \\
$\nu_{eff}$ $\geq$1.58&5&416380&-7.3$\pm$1.3 & -10.0$\pm$1.0 & 5.9$\pm$1.4 &-8.6$\pm$1.4  &4.0$\pm$1.1  &12.0$\pm$1.1 \\
$\nu_{eff}$<1.58&5&140489&6.9$\pm$2.5 & 10.6$\pm$2.5 & 5.5$\pm$3.0 &-15.1$\pm$2.8  &7.2$\pm$2.8  &13.4$\pm$2.5 \\
\hline
& &  &  &  &&  &  &    \\
KQCG & 1 &208743&9.6$\pm$2.6  &7.6$\pm$1.9  &-16.3$\pm$2.6  &-  &- &-   \\
 $\nu_{eff}$ $\geq$1.58& 1 & 185360&7.8$\pm$2.7  &6.7$\pm$2.0  &-16.7$\pm$2.7  &-  &- &- \\
$\nu_{eff}$<1.58&1&22526 &25.5$\pm$8.1  &18.3$\pm$6.4&-10.7$\pm$8.3  &-  & - &-\\
\hline
& &  &  &  &&  &  &    \\
\multirow{3}{*}{KQCG+GCRF2} & 1 &765612 &-2.2$\pm$0.8 & -1.2$\pm$0.7 & -2.0$\pm$0.8 & -6.3$\pm$0.8 & 4.7$\pm$0.7 & 11.8$\pm$0.7 \\
 & 5 &765612& -4.5$\pm$1.1 &-6.8$\pm$0.9 & 5.2$\pm$1.2 & -9.1$\pm$1.2 & 4.5$\pm$1.0 & 11.7$\pm$1.0 \\
 & 10 &765612& -4.6$\pm$1.5 &-7.6$\pm$1.2 & 6.7$\pm$1.5 & -11.7$\pm$1.6 & 5.0$\pm$1.2 & 13.2$\pm$1.3 \\
$\ast$&5&765612&-4.5$\pm$1.1&-6.8$\pm$0.9&6.5$\pm$1.2&-9.0$\pm$1.2&4.6$\pm$1.0&-1.5$\pm$0.9\\
& &  &  &&  &  &  &    \\
\multirow{1}{*}{19$\leq$G<20}  & 5 & 329900 &11.2$\pm$2.2  &1.4$\pm$1.7  &6.6$\pm$2.4  &-8.4$\pm$2.4  &1.8$\pm$1.9& 15.7$\pm$1.9\\
\multirow{1}{*}{G$\geq$20}  & 5 &273836 &5.1$\pm$5.6  &23.3$\pm$4.6  &9.0$\pm$6.1  &-15.6$\pm$6.2  &5.4$\pm$4.8  &6.7$\pm$4.9 \\
\hline
& &  &  &  &&  &  &    \\
\multirow{2}{*}{Type3} & 1 &485985&-3.8$\pm$0.8  &-3.2$\pm$0.7  &-0.9$\pm$0.9  &-6.9$\pm$0.8  &4.6$\pm$0.8 &11.5$\pm$0.8   \\
 & 5 & 485985&-5.0$\pm$1.1  &-8.4$\pm$0.9  &5.6$\pm$1.2  &-10.0$\pm$1.2  &4.9$\pm$1.0 &10.8$\pm$1.0 \\
$\ast$&5&485985 &-5.0$\pm$1.1  &-8.4$\pm$0.9&5.2$\pm$1.2  &-9.9$\pm$1.2  & 5.0$\pm$1.0 & -3.3$\pm$1.0\\
&& &  &&  &  &  &    \\
\multirow{2}{*}{Type2} & 1 &2843& -25.0$\pm$6.2 &-1.5$\pm$5.8 & 2.0$\pm$6.6 & -8.8$\pm$6.3 & -1.1$\pm$6.2 & 24.7$\pm$5.5 \\
 & 5 &2843& -28.1$\pm$7.8 & -2.8$\pm$7.1 & 5.0$\pm$8.6 & -9.1$\pm$8.7 & 8.0$\pm$7.6 & 20.0$\pm$7.0 \\
$\ast$&5&2843 &-28.0$\pm$7.8  &-2.9$\pm$7.1&-14.1$\pm$8.6  &-9.0$\pm$8.7  & 7.8$\pm$7.6 & -2.8$\pm$7.0\\
\hline
\end{tabular}
\end{table*}

\begin{table*}
\centering
\caption{Global rotation of different quasar subsets in proper motions. All solutions are weighted.}
\label{rotation_result}
\begin{tabular}{ccccc}
\hline
\hline
Subset & N& \begin{tabular}[c]{@{}c@{}}$w_{X}$ \\ ($\mu$as/yr)\end{tabular} & \begin{tabular}[c]{@{}c@{}}$w_{Y}$\\ ($\mu$as/yr)\end{tabular} & \begin{tabular}[c]{@{}c@{}}$w_{Z}$\\ ($\mu$as/yr)\end{tabular} \\
\hline
&&&\\
KQCG+GCRF2 &765612&  -2.1$\pm$0.8&-0.8$\pm$0.7& -2.4$\pm$0.8         \\
North      & 465093  &-3.4$\pm$1.0       & -2.2$\pm$0.9  & -7.3$\pm$1.2 \\
South      & 300519  &  0.0$\pm$1.1                & 0.9$\pm$1.0                & 3.0$ \pm$1.2 \\
\hline
&&&&\\
GCRF2 &556869&  -3.1$\pm$0.8&-1.9$\pm$0.7& -1.0$\pm$0.9  \\
North      & 285806  &-5.6$\pm$1.1       & -4.5$\pm$1.0  & -5.2$\pm$1.3   \\
South      & 271063  &  -0.3$\pm$1.2                & 0.8$\pm$1.0                & 3.0$ \pm$1.2     \\
\hline
&&&&\\
Type3  &485985 &-3.3$\pm$0.8& -2.8$\pm$0.7  & -0.9$\pm$0.9      \\
North        &247999& -5.7$\pm$1.1& -5.5$\pm$1.0 & -5.1$\pm$1.3 \\
South         &237986&  -0.6$\pm$1.2 & -0.1$\pm$1.0              & 2.9$ \pm$1.3 \\
\hline
&&&&\\
Type2    &2843&  -23.1$\pm$5.8 & 2.3$\pm$5.4  & 2.7$\pm$5.6  \\
North  & 1635&-25.1$\pm$7.1 & -6.9$\pm$6.5  & 2.6$\pm$8.5    \\
South   & 1208& -17.8$\pm$10.3& 9.0$\pm$9.7 & 1.5$ \pm$10.7   \\
\hline
\end{tabular}
\end{table*}

We also tried to fit a pure rotation to the proper motions, using the following equations \citep{Mignard2012}:

\begin{equation}
\begin{array}{l}
\mu_{\alpha\ast} = -w_{X}\cos\alpha\sin\delta-w_{Y}\sin\alpha\sin\delta+w_{Z}\cos\delta\\
\mu_{\delta} = +w_{X}\sin\alpha-w_{Y}\cos\alpha
\end{array}
\label{rotation}
\end{equation}

Where $w_{X}$, $w_{Y}$, and $w_{Z}$ are the three spin rates of the proper motion field. We apply this fit to further investigate the spin rate of different quasar subsets in the northern and southern hemisphere. The results are shown in Table \ref{rotation_result}.  For the Type2 quasars, no significant spin difference between the two hemispheres is found. However, for the other quasar subsets, the spin rate is clearly above the statistical noise in the northern hemisphere, but negligible in the southern one;  this feature could be explained by a north/south dichotomy  in the magnitude and color distribution of the fitted quasars, or by a global positional rotation between the northern and southern subsets inducing a rotation in the proper motion field.

\section{The scalar spherical harmonics expansion of parallaxes}
The parallaxes of quasars can be treated as parallax residuals, and can be seen as the radial part of spatial position differences on the celestial sphere. Therefore, they represent a scalar field on the sphere surface that can be expanded in terms of spherical harmonics (SSH) as follows \citep{bucciarelli2011}:

\begin{equation}
\Delta\pi=V_{\pi}(\alpha,\delta)=\sum_{l}\sum_{m=-l}^{l}c_{lm}Y_{lm}(\alpha,\delta)
\label{ssh_1}
\end{equation}
Where $Y_{lm}$ are the standard spherical functions defined here by the following sign convention:

\begin{equation}
Y_{lm}=(-1)^m\sqrt{\frac{2l+1}{4\pi}\frac{(l-m)!}{(l+m)!}}P_{lm}(\sin\delta)e^{im\alpha}
\label{ylm}
\end{equation}

for $m\geq0$, and we have $Y_{l,-m}(\alpha,\delta)=(-1)^mY_{lm}^{\ast}(\alpha,\delta)$ for $m <0$. The ${\ast}$ denotes complex conjugation, and  $P_{lm}(x)$ are the associated Legendre polynomials.

Equation \ref{ssh_1} can be reduced as:
\begin{equation}
\Delta \pi(\alpha,\delta)=\sum^{l_{max}}_{l=1}\left[c^R_{l0}Y^R_{l0}+2\sum^{l}_{m=1}\left(c^R_{lm}Y^R_{lm}-c^I_{lm}Y^I_{lm}\right)\right]
\label{ssh}
\end{equation}
Where $R$ and $I$ denote the real and imaginary part of the function.
Starting from the definition of $power$ as the integral of the squared function divided by the domain area,  by virtue of  Parseval's Theorem we can express the total $power$ per degree $l$ of  $\Delta\pi(\alpha,\delta)$  in terms of the expansion coefficents as  
\begin{equation}
P_l=(c^R_{l0})^2+2\sum^l_{m=1}\left[(c^R_{lm})^2+(c^I_{lm})^2\right]
\label{pow}
\end{equation} 
Normalizing each coefficient of the above sum by  its formal error, assuming white Gaussian noise, we obtain a $\chi^2$-distributed variable with $2l+1$ degrees of freedom, which can be used to test the statistical significance of the corresponding degree. A more robust form of test variable, still $\chi^2$-distributed, is given by equation (87) of \citet*{Mignard2012},  or the derived quantity $Z_{\chi^2}$ which follows a standard normal distribution (see Eq. (85) of \cite{Mignard2012}) and it  is the one we used in the present  analysis. The results of the SSH analysis, having subtracted  beforehand the bias to each parallax, are summarized in Table \ref{ssh}. Note that a value of $Z_{\chi^2}> 2.33$ corresponds to a confidence level of $99\%$, or $2.33\sigma$ of a normal distribution. The parameter $(P_l/4\pi)^{1/2}$ represents the RMS value of the scalar field for the corresponding degree $l$.  The expansion of Type2 subset does not present particular signatures, while the other subsets show significant powers for degrees $l=1$ and $l=4$. The total RMS value for $l\leq 10$ (angular scales $\ge$ $180/l$=18 degree) of each subset is about 13 $\mu as$ (apart from Type2). Using a different spatial correlation technique,  the Gaia-DR2 Astrometry paper \citep{Lindegren2018Gaia} reports an angular scale of 14 degrees with a RMS amplitude of 17 $\mu as$, which  is in good agreement with our results.

\begin{table*}
\centering
\caption{The spherical harmonics expansion of the parallaxes of different quasar subsets. The parallax bias is subtracted before the expansion. All solutions are weighted.}
\label{ssh}
\begin{tabular}{ccccccccccc}
\hline
\hline
&&&&\\
& \multicolumn{2}{c}{KQCG+GCRF2}            & \multicolumn{2}{c}{Type2+Type3}   & \multicolumn{2}{c}{Type3} & \multicolumn{2}{c}{Type2} & \multicolumn{2}{c}{GCRF2}                 \\
\hline
&&&&&&&&\\
    $l$ & $(P_l/4\pi)^{1/2}(\mu as)$ & $Z_{\chi^2}$ & $(P_l/4\pi)^{1/2}(\mu as)$ & $Z_{\chi^2}$ & $(P_l/4\pi)^{1/2}(\mu as)$ & $Z_{\chi^2}$ & $(P_l/4\pi)^{1/2}(\mu as)$ & $Z_{\chi^2}$ & $(P_l/4\pi)^{1/2}(\mu as)$ & $Z_{\chi^2}$ \\
\hline
&&&&&&&&\\
1                  & 5.1                  & 5.9              & 5.9                  & 7.2              &5.8&7.0&12.3&1.2&5.2&5.9 \\
2                  & 3.1                  & 2.6               & 2.8                  & 2.4               &3.1&2.7&18.7&2.1&2.8&2.0\\
3                  & 4.2                  & 4.2               & 4.9                  & 5.5               &4.8&5.3&16.7&1.1&4.3&4.2\\
4                  & 5.5                  & 5.6               & 6.8                  & 7.8             &7.0&7.9&16.8&0.6 &5.9&6.0 \\
5                  & 4.9                  & 5.0               & 4.6                  & 4.7              &4.6&4.6&21.2&1.5&4.8&4.7\\
6                  & 3.3                  & 1.8              & 4.1                  & 3.5               &4.4&3.9&24.7&2.1&3.3&1.7\\
7                  & 4.2                  & 3.9               & 4.2                  & 4.0               &4.3&4.0&21.7&1.1&4.0&3.4\\
8                  & 3.3                  & 1.8               & 3.2                  & 1.6               &3.3&1.8&18.5&-0.2&3.2&1.5\\
9                  & 3.3                  & 2.7               & 3.5                  & 3.1              &3.5&2.9&24.5&1.8 &3.5&2.9\\
10                 & 3.9                  & 3.8               & 3.4                  & 2.6             &3.4&2.5&27.0&2.6&3.8&3.3 \\
\hline
\end{tabular}
\end{table*}

\section{ICRF2 sources in Gaia DR2}
In this section, we compare the VLBI positions of ICRF2 sources \citep{Fey2015The} with their optical counterparts in Gaia DR2.

\begin{figure}
   \centering
   \includegraphics[width=6cm]{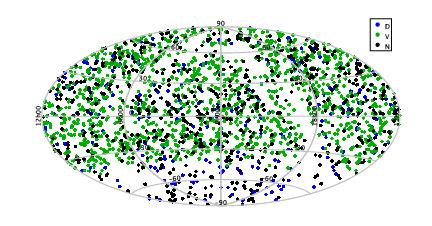}
      \caption{Sky distribution of ICRF2 sources found in Gaia DR2, Hammer-Aitoff projection in equatorial coordinates. Blue dots are defining sources (D), green dots are VLBA Calibrator Survey sources (VCS  V), and  blacks are non VCS sources (N).}
         \label{FigICRFDR2skydensity}
   \end{figure}

\begin{figure}
   \centering
   \includegraphics[width=5.5cm]{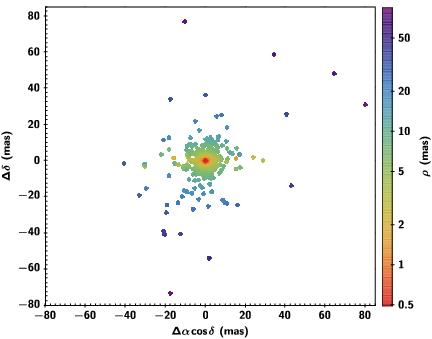}
      \caption{Scatter plot of position differences in right ascension and declination (Gaia DR2 minus ICRF2). }
         \label{PD_H}
   \end{figure}

After cross-matching, 2146 ICRF2 sources are found in the Gaia DR2 sources, with sky distribution given in Figure \ref{FigICRFDR2skydensity}. Most angular differences $\rho$ between matched sources are smaller than 1 mas, and just a few sources have $\rho>10$ mas, see  Figure \ref{PD_H} for color-coded scatter plot of position differences $\rho$ in right ascension and declination.

   \begin{figure}
   \centering
   \includegraphics[width=6cm]{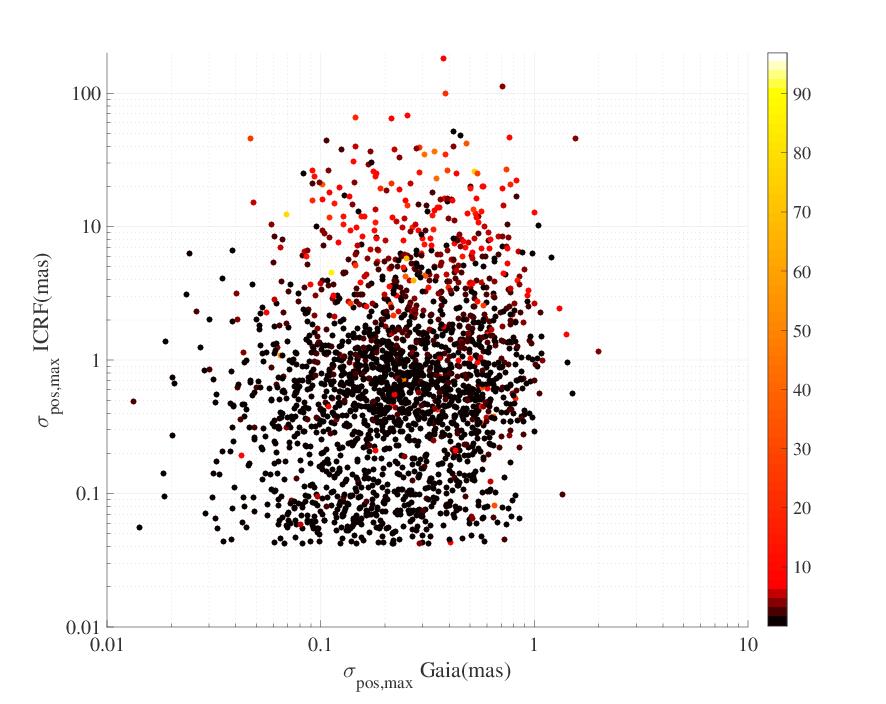}
      \caption{The formal position uncertainties $\sigma_{pos,max}$ of the Gaia DR2 sources (abscissa) with respect to the ICRF2 sources  (ordinate). The color bar on the right is the position differences $\rho$ (in mas) between Gaia DR2 and the ICRF2 sources. The axis is log-log scale.}
         \label{rhoVsig}
   \end{figure}

Figure \ref{rhoVsig} shows a plot of color-coded angular separations between matched sources in the plane of formal positional uncertainties $\sigma_{DR2}$ , $\sigma_{ICRF2}$ . Most of the sources in Gaia DR2 have position uncertainties under 1 mas, while the uncertainties of the sources in ICRF2  range from 0.04 mas to 10 mas with few even up to tens of mas. Some sources with small position uncertainties show large angular differences, which may caused by an offset between the centers of emission at optical and radio wavelengths.

The alignment of the optical positions in Gaia DR2 with respect to the ICRF2 can be modelled by an infinitesimal solid rotation with the following equations \citep{Mignard2012}:
\begin{equation}
\begin{array}{l}
\Delta\alpha_{\ast}=-\epsilon_{X}\cos\alpha\sin\delta-\epsilon_{Y}\sin\alpha\sin\delta+\epsilon_{Z}\cos\delta\\
\Delta\delta=+\epsilon_{X}\sin\alpha-\epsilon_{Y}\cos\alpha
\end{array}
\label{glodiff}
\end{equation}
Where $\Delta\alpha_{\ast}=\Delta\alpha\cos\delta$, and $\epsilon_{X}$, $\epsilon_{Y}$ and $\epsilon_{Z}$ are the three rotation angles between the two reference frames.

\begin{table*}
\centering
      \caption{Global difference between the Gaia-CRF2 positions of ICRF sources and their positions in ICRF2. }
         \label{globaldiff}
\begin{tabular}{ccccc}
\hline
\hline
& &  &  &  \\
 Subset & N & $\epsilon_{X}$ ($\mu$as) & $\epsilon_{Y}$ ($\mu$as) & $\epsilon_{Z}$ ($\mu$as)\\
\hline
& &  &  &  \\
All&2146&-3.6$\pm$ 27.5&27.2$\pm$26.9  &3.8$\pm$25.7\\
 Defining&257&-19.1 $\pm$ 36.2&30.4$\pm$ 35.0&-32.9$\pm$ 37.0\\
 Non-defining&1889&12.1 $\pm$ 37.7&25.2 $\pm$ 37.1&26.6$ \pm$ 32.9\\
\hline
\end{tabular}
\end{table*}

The weighted least-squares estimation of the orientation parameters between Gaia-CRF2 and ICRF2 are listed in table \ref{globaldiff}. No significant rotation is found at the level of 0.03 $mas$ in position. This indicates that the axes of Gaia-CRF2 and the ICRF2 are well aligned with each other within 30 $\mu as$.

\section{Conclusions}
We cross-matched the quasars from the compilation of the SDSS-DR14, LQAC3 and LAMOST DR5 with Gaia DR2, and found 208743 extra quasars in Gaia DR2, which is about $37\%$ of the Gaia-CRF2 sample. We used  this independent sample and the already known quasars in DR2 to investigate the properties of the QSOs solution,  also by comparing  the astrometric residuals of various quasar subsets in DR2.  In general, we obtained consistent results between the samples;  some signatures varying with different subsets, and clearly above the statistical noise, are still compatible with systematic errors depending on source position, magnitude and color not completely cured in the second release of the Gaia data, as discussed in the Gaia-DR2 astrometry paper. 
The results of our analysis are summarized below :
   \begin{enumerate}
      \item The parallaxes of our KQCG sample have a mean bias of  $-0.0330$ $mas$ and a median of $-0.0278$ $mas$, which agree well with the results of the GCRF2 sample; we note, however,  that the mean parallax of Type2 subset in GCRF2 is $0.02$ $mas$ smaller. 
\item There is a $-9.1$ $\mu as/yr$ bias in $\mu_{\alpha\ast}$ of the KQCG sample, and a bias of about $+10$ $\mu as/yr$ in $\mu_{\delta}$ for all quasar subsets. The mean systematic error in $\mu_{\alpha\ast}$ trends from positive to negative at the effective wavenumber $\nu_{eff}\sim$ 1.58 $\mu m^{-1}$ for  the GCRF2 sample.
      \item The VSH method is applied to the proper motion vector field of different quasar subsets. The results for different subsets agree with each other. For Type2, no significant rotation difference between northern and southern hemisphere is found. However, the GCRF2 and other subsets shows a different rotation between two hemispheres.
      \item The spherical harmonics expansion of the parallaxes shows an angular scale of 18 deg with an RMS amplitude of 13 $\mu as$.
\item The comparison of the VLBI-based positions of ICRF2 sources and their Gaia DR2 counterparts shows that the axes of Gaia-CRF2 and the ICRF2 are well aligned with each other within 30 $\mu as$. 
   \end{enumerate}

%
%
%
%
%
%
%
%
%
%
%
%
%
%

\section*{Acknowledgements}

This work has made used of data from ESA space mission Gaia, processed by the Gaia Data Processing and Analysis Consortium (DPAC). We are grateful to the developers of the TOPCAT (\citep{Taylor2005TOPCAT}) software. This work has been supported by the grants from the National Science Foundation of China (NSFC) through grants 11703065, 11573054 and 11503042.




\bibliographystyle{mnras}
\bibliography{mnras_ref.bib} 




%
%


\bsp	
\label{lastpage}
\end{document}